\documentclass[a4paper]{cas-sc}

\usepackage[authoryear,longnamesfirst]{natbib}

\usepackage{amsmath}
\usepackage{indentfirst}
\usepackage{amsfonts}
\usepackage{mathrsfs}
\usepackage{amssymb}
\usepackage{geometry}
\usepackage{latexsym}
\usepackage[all]{xy}

\usepackage{amsthm}

\def\tsc#1{\csdef{#1}{\textsc{\lowercase{#1}}\xspace}}
\tsc{WGM}
\tsc{QE}

\date{}

\begin{document}
	
\let\WriteBookmarks\relax
\def\floatpagepagefraction{1}
\def\textpagefraction{.001}



\title [mode = title]{The characterization of hyper-bent function with multiple trace terms in the extension field}

\tnotemark[1]


\tnotetext[1]{This work was supported by the National Natural Science Foundation of China under Grant No.12361001.}

%
\author[1]{Peng Han }[style=chinese, orcid=0000-0002-4753-7206]                           



\ead{penghanmitp@163.com}



\affiliation[1]{ organization={School of Mathematics},addressline={Aba Teachers University},city={Wenchuan},state={Sichuan},postcode={623000},country={China}       }

\author[1,2]{Keli Pu}[style=chinese]
\cormark[1]


\ead{pp180896@163.com}



\affiliation[2]{ organization={School of Mathematical and  Sciences},addressline={Sichuan Normal University},city={Chengdu},state={Sichuan},postcode={610066},country={China}       }            




    
\begin{abstract}
Bent functions are maximally nonlinear Boolean functions with an even number of variables, which include a subclass of functions, the so-called hyper-bent functions whose properties are stronger than bent functions and a complete classification of hyper-bent functions is elusive and inavailable.~In this paper,~we solve an open problem of Mesnager that describes hyper-bentness of hyper-bent functions with multiple trace terms via Dillon-like exponents with coefficients in the extension field~$\mathbb{F}_{2^{2m}}$~of this field~$\mathbb{F}_{2^{m}}$. By applying M\"{o}bius transformation and the theorems of hyperelliptic curves, hyper-bentness of these functions are successfully characterized in this field~$\mathbb{F}_{2^{2m}}$ with~$m$~odd integer.  
\end{abstract}






\maketitle

\section{Introduction}\label{1}
\indent Boolean functions play a vital role in cryptographic applications in terms of its nonlinearity based criterion, such as design of substitution box in block cipher,~generation of pseudo-random key stream in stream cipher.~Let~$n=2m$~be an even positive integer number.~Let~$\mathbb{F}_{2^{n}}$~be the finite field with~$2^{n}$~elements and~$\mathbb{F}^{\ast}_{2^{n}}$~be the multiplication group of~$\mathbb{F}_{2^{n}}$, where items are not including~0~element over the finite field~$\mathbb{F}_{2^{n}}$.~Let~$f$~be a Boolean function over~$\mathbb{F}_{2^{n}}$,~$\chi_{f}(x)=(-1)^{f(x)}$~is its sign function.~For~$x,y\in\mathbb{F}_{2^{n}}$,~the inner product is written as~$x\cdot y=\mathrm{Tr}^{n}_{1}(xy)$.~The~Walsh Hadamard~transform of~$f$~is the discrete Fourier transform of~$\chi_{f}(x)$,~i.e.,
\begin{align*}
\centering	\widehat{\chi}_{f}(\omega)=\sum\limits_{x\in\mathbb{F}_{2^{n}}}(-1)^{    f(x)+\mathrm{Tr}^{n}_{1}(\omega x) },~\forall \omega\in\mathbb{F}_{2^{n}}.
\end{align*}
\noindent In addition,~the Walsh Hadamard~transform~satisfies~Parseval~relation
\[\sum\limits_{\omega\in\mathbb{F}_{2^{n}}} {\widehat{\chi}_{f}}^{2}(\omega)= 2^{2n}    .       \]
\indent Bent functions in 1976 introduced by Rothaus~[1]~have attracted more and more attention with achieving the highest possible nonlinearity and only exist for the Boolean functions of even numbers of variables by Parseval equation and have been extensively applied in the fields of coding theory~[2, 3], cryptography~[4], sequence designs~[5], graph theory~[6, 7], and association scheme~[8].~The Walsh Hadamard~transform is a powerful tool to determine bentness of Boolean function.~A Boolean function in~$\mathbb{F}_{2^{n}}$~is said to be bent if and only if~$\widehat{\chi}_{f}(\omega)=\pm 2^{\frac{n}{2}}$.~Up to now,~a complete classification of bent functions is still vague.~However, a number of interesting results on bent functions have been found via primary constructions and secondary constructions~[9, 10, 11, 12, 13, 14, 15, 16, 17, 18, 19, 20, 21, 22].~Hyper-bent function is a subclass of bent functions, which possesses even stronger properties than bent functions were introduced by Youssef and Gong~[23]~in 2001.~In fact, the initial definition of hyper-bent functions was proposed by Golomb and Gong~[24]~in 1999.~The extended~Walsh Hadamard~transform of~$f$~is defined by the following form:
\[ \widehat{\chi}_{f}(\omega,k)=\sum\limits_{x\in\mathbb{F}_{2^{n}}}(-1)^{ f(x)+\mathrm{Tr}^{n}_{1}(\omega x^{k}) },~\forall \omega\in\mathbb{F}_{2^{n}},\]
where~$gcd(k,2^{n}-1)=1$.~By employing the extended Walsh Hadamard~transform,~the concept of hyper-bent functions can be precisely defined,~i.e.,~a bent function~$f$~is said to be the hyper-bent function if and only if~$\widehat{\chi}_{f}(\omega,k)=\pm 2^{\frac{n}{2}}$.~The algebraic degree of bent function is at most~$n/2$.~Specificly, hyper-bent function in~$\mathbb{F}_{2^{n}}$~is exact~$n/2$.~The classifications of hyper-bent functions seem fuzzy.~Monomial hyper-bent functions in~$\mathbb{F}_{2^{n}}$~are famous bent functions discovered by Dillon~[13].~Charpin and Gong~[25]~have characterized a large class of hyper-bent functions involving Dickson polynomials, which includes the well-known monomial functions with Dillon exponent~$r(2^{m}-1)$,~where~$r$~is a positive integer.~By applying Kloosterman sums, we can determine a bent function whether it is a hyper-bent function.
Let~$a$~be an element over the finite field~$\mathbb{F}^{\ast}_{2^{m}}$.~The class of monomial functions~$\mathrm{Tr}^{n}_{1}(ax^{r(2^m-1)})$~is hyper-bent if and only if the value of Kloosterman sums satisfies~$K_{m}(a)=0$ in~[26, 27, 28].~A family of binomial functions is captured with~$f^{(r)}_{a,b}(x)=\mathrm{Tr}^{n}_{1}(ax^{r(2^{m}-1)})+\mathrm{Tr}^{2}_{1}(bx^{\frac{ 2^{n}-1}{3}})$,~where~$a\in\mathbb{F}^{\ast}_{2^{n}}$~and~$b\in\mathbb{F}^{\ast}_{4}$.~According to Mesnager's~[29],~when~$r=1$,~it is a hyper-bent function if and only if the value of Kloosterman sums on coefficient~$a$~is 4.~At the same time,~Mesnager professor in~[30]~also presents the fact that the value of Kloosterman sums satisfies~$K_{m}(a)=4$~if~$f^{(1)}_{a,b}(x)$~is a bent functions,~where~$m$~is an even positive integer.~There is an open problem~[31]~left by Mesnager in~[29]~for the binomial bent functions with the case~$m$~even.~Recently, this problem is resolved by Tang et al.~[32]~applying computing tools, which include Walsh transform of Boolean functions by means of multiplicative characters, Gauss sums and relative properties of graph theory, i.e., if~$K_{m}(a^{2^{m}+1})=4$, then~$f^{(1)}_{a,b}(x)$~is bent function.~Let~$D_{r}(x)$~be~$r$-th order Dickson polynomial.~Charpin and Gong in~[25]~studied the hyper-bentness of Boolean functions of this form~$f_{a_{r}}(x)=\sum\limits_{r\in R}\mathrm{Tr}^{n}_{1}(a_{r}x^{r(2^m-1)})$,~where~$R$~consists of a set of representatives of the cyclotomic coset modulo~$2^{m}+1$~and the coefficient~$a_{r}$~lies in this field~$\mathbb{F}_{2^m}$.~Charpin and Gong prove the fact if all of~$a_{r}$~belong to subfield~$\mathbb{F}_{2^m}$,~then~$f_{a_{r}}(x)$~is a hyper-bent function if and only if~$\sum\limits_{x\in\mathbb{F}_{2^m}}(-1)^{\mathrm{Tr}^{m}_{1}(x^{-1})+g(x) }=2^m-2wt(g)$,~where~$g(x)=\sum\limits_{ r\in R}\mathrm{Tr}^{m}_{1}(a_{r}D_{r}(x)) $.~\\
\indent~Mesnager~[33, 34]~has characterized another class of hyper-bent functions,~which is distinct from that of Charpin and Gong by means of Dickson polynomials.~This function~$f_{a_{r},b}(x)=\sum\limits_{ r\in R}\mathrm{Tr}^{n}_{1}(a_{r}x^{r(2^{m}-1)})+\mathrm{Tr}^{2}_{1}(bx^{\frac{ 2^{n}-1}{3}})$~is extended by Mesnager to deal with Charpin-Gong like functions,~where the same restriction coefficient~$a_{r}$~lives in~$\mathbb{F}_{2^{m}}$~and~$b$~is in~$\mathbb{F}_{4}$.~In addition,~the coefficient~$b$~is in~$\mathbb{F}_{16}$~and~$m\equiv2(mod\,4)$,~the hyper-bentness of this function~$f_{a_{r},b}(x)=\sum\limits_{ r\in R}\mathrm{Tr}^{n}_{1}(a_{r}x^{r(2^{m}-1)})+\mathrm{Tr}^{4}_{1}(bx^{\frac{ 2^{n}-1}{5}})$~was discussed in~[35].~We generalize the results in~[34]~of the hyper-bentness of Charpin-Gong like functions with an additional trace term over the finite field~$\mathbb{F}_{4}$~by the form~$f^{(r)}_{a_{r},b}(x)=\sum\limits_{ r\in R}\mathrm{Tr}^{n}_{1}(a_{r}x^{r(2^{m}-1)})+\mathrm{Tr}^{2}_{1}(bx^{\frac{ 2^{n}-1}{3}})$, where the coefficient~$a_{r}$~lies in the extension field~$\mathbb{F}_{2^{n}}$~rather than in~$\mathbb{F}_{2^{m}}$.~\\
\indent~The rest parts of this paper are organized as follows. In Section 2,~some elementary notations and preliminaries are recalled. In Section 3,~the hyper-bentness of~$f^{(r)}_{a_{r},b}(x)$~is presented with coefficient~$a_{r}$~living in~$\mathbb{F}_{2^{n}}$.~In Section 4,~we conclude the relative results in this paper.










\section{Preliminaries}\label{2}
\indent Throughout this paper,~let~$n=2m$~be an even positive integer,~where~$m$~odd.~Let~$\mathbb{F}_{q}$~be a finite field with~$q$~elements. While working over the finite field~$\mathbb{F}_{2^{n}}$,~we denote the notation~$\frac{1}{0}=0^{2^{n}-2}=0$.~The set~$U_{2^{m}+1}$~is always the group of~$(2^{m}+1)$-th~roots of unity that is~$U_{2^{m}+1}=\{u\in\mathbb{F}^{\ast}_{2^n}|u^{2^{m}+1}=1\}$.
\subsection{Trace functions and Hyper-bent functions}\label{2.1}
\indent Trace function plays a crucial role in cryptography and coding theory.~For a positive integer~$k$~dividing~$n$.~Trace function is a map from~$\mathbb{F}_{2^{n}}$~to~$\mathbb{F}_{2^{k}}$~denoted by~$ \mathrm{Tr}^{n}_{k}(\cdot)$.~It is defined as
\[ \mathrm{Tr}^{n}_{k}(x)=x+x^{2^{k}}+\cdots+x^{2^{n-k}} , \forall x\in\mathbb{F}_{2^{n}}.    \]
In particular,~$\mathrm{Tr}^{n}_{1}(x)$~is called the absolute trace function and has the following transitivity property:
\[  \mathrm{Tr}^{n}_{1}(x) =  \mathrm{Tr}^{k}_{1}(  \mathrm{Tr}^{n}_{k}(x) ).      \]
\indent~Let~$\Gamma_{n}$~be the set of integers obtained by selecting an element in each cyclotomic coset modulo~$2^{n}-1$.~$o(j)$~is the smallest positive integer containing~$j$~such that~$j2^{o(j)}\equiv j(2^{n}-1)$~holds.~Every non-zero Boolean function in~$\mathbb{F}_{2^{n}}$~can be expressed as the following form:
\[ f(x)=\sum\limits_{j\in\Gamma_{n}}\mathrm{Tr}^{o(j)}_{1}(a_{j}x^{j})+\varepsilon(1+x^{2^{n}-1}), \]
where~$a_{j}\in\mathbb{F}_{2^{o(j)}}$,~$\varepsilon= wt(f)(mod\,2)$~and~$wt(f)$~is the cardinality of support set of~$f$,~i.e.,~$\# supp=\{x\in\mathbb{F}_{2^{n}}\mid f(x)=1\}$.~\\
\indent~Let~$\alpha $~be a primitive element of~$\mathbb{F}_{2^{n}}$.~Let~$f$~be a Boolean function of~$\mathbb{F}_{2^{n}}$~such that~$f(\alpha^{2^{m}+1}x)=f(x)$~and~$f(0)=0$. This function~$f$~is a hyper-bent function if and only if the weight of the vector~$(f(1),f(\alpha),f(\alpha^{2}),\cdots,f(\alpha^{2^{m}}) )$~
equals~$2^{m-1}$.~\\
~Let~$f:\mathbb{F}_{2^{m}} \rightarrow \mathbb{F}_{2}$~and~$u\in U_{2^{m}+1}$,~the exponential sum~$\Lambda(f)$~in~$\mathbb{F}_{2^n}$~is defined as
\begin{align*}~\tag{1}
\Lambda(f)=\sum\limits_{u\in U_{2^{m}+1}}(-1)^{f(u)},
\end{align*}
and~$f$~is hyper-bent if and only if~$\Lambda(f)=1$.~It is a perfect tool to study the hyper-bentness of Boolean functions~$f$~of the polynomial forms.
\subsection{M\"{o}bius transformations}\label{2.2}
A M\"{o}bius transformation is also called the fractional linear mapping that is a rational function of the form
\[  ~\frac{ax+b}{cx+d},~~a,b,c,d\in\mathbb{F}_{2^{m}}~and~ ad+bc\neq 0  ,      \]
where variable~$x\in\mathbb{F}_{2^{m}}\cup{\{\infty\}}$.~Fix~$u_{0}\in U_{2^{m}+1}\setminus{\{1\}}$.~By using M\"{o}bius transformation,~we can present the elements of~$U_{2^{m}+1}$~as
\begin{align*}~\tag{2}
U_{2^{m}+1}=\left \{ \frac{(u_{0}+1 )x+1} {(u_{0}+1 )x+u_{0}} : x\in\mathbb{F}_{2^{m}}\cup{\{\infty\}}    \right \}
\end{align*}
Specially,~we select the element~$u$~of~$U_{2^{m}+1}$.~If~$x=\infty$,~then~$u=1$.~Assume that~$\rho_{0}=\frac{u_{0}}{1+u^{2}_{0}}$,~we get the results
\begin{align*}~\tag{3}
u+\frac{1}{u} =\frac{1}{x^{2}+x+\rho_{0}}
\end{align*}
and
\begin{align*}~\tag{4}
u_{0}u+\frac{1}{u_{0}u} =\frac{x^{2}}{\rho_{0}(x^{2}+x+\rho_{0})}  .
\end{align*}
They play a vital role for us to research the hyper-bentness of Boolean funtions in the extension fields~$\mathbb{F}_{2^{2m}}$~of finite field~$\mathbb{F}_{2^{m}}$.
\subsection{Dickson polynomial }\label{2.3}
Dickson~polynomial of~degree~$r$~in~$\mathbb{F}_{2}[X]$~is defined as
\[ D_{r}(X)=\sum\limits^{\lfloor{\frac{r}{2}}\rfloor}_{i=0}\frac{r}{r-i}C^{i}_{r-i}X^{r-2i} ,r=2,3,\cdots                       \]
where~$\lfloor{  r/2}\rfloor$~denotes the largest integer less than or equal to~$r/2$. It has many applications and interesting properties~[36].~Dickson polynomials~$D_{r}(X)$~satisfy the following recurrence relation
\[ D_{i+2}(X)=XD_{i+1}(X)+D_{i}(X) ,                    \]
the initial value of Dickson polynomial~$D_{0}(X)=0$,~$D_{1}(X)=X$.~For any non-zero positive integer~$r$~and~$p$,~there are the following properties about Dickson polynomials~\\
\[     D_{rp}(X)=D_{r}(D_{p}(X)).                \]
Dickson polynomials satisfy with~$X=1/x$~and~$x\neq 0$,
\[ D_{r}(x+x^{-1})=x^{r}+x^{-r}.  \]
\section{The characterization of hyper-bent function in the extension field}\label{3}
\indent Recently,~Claude Carlet et al.~[37]~have researched the hyper-bentness of Boolean functions in trace representation given by Youssef and Gong~[23]~in the following form:
\begin{align*}~\tag{5}
f_{a_{r}}(x)=\sum\limits_{ r\in R}\mathrm{Tr}^{n}_{1}   (a_{r}x^{r(2^m-1)}) ,
\end{align*}
where~$R$~is a set of representatives of the cyclotomic classes modulo~$2^{m}+1$,~and the coefficient~$a_{r}$~lives in this field~$\mathbb{F}_{2^{n}}$~rather than its subfield~$\mathbb{F}_{2^{m}}$.~In this section, we use the technology of Claude Carlet et al. to treat the hyper-bentness of Charpin-Gong like functions with an additional trace term over~$\mathbb{F}_{4}$:~\\
\begin{align*}~\tag{6}
f^{(r)}_{a_{r},b}(x)=\sum\limits_{ r\in R}\mathrm{Tr}^{n}_{1}(a_{r}x^{r(2^m-1)})+\mathrm{Tr}^{2}_{1}(bx^\frac{2^n-1}{3}),
\end{align*}
where the coefficient~$b$~is in~$\mathbb{F}_{4}$, and the coefficient~$a_{r}$~lies in this field~$\mathbb{F}_{2^{n}}$.~\\

\noindent{\bf Theorem 3.1}
Let~$u_{0}$~be a primitive~$(2^{m}+1)$-th~root of unity and~$\rho_{0}=\frac{u_{0}}{1+u^{2}_{0}}$.~Let~$f^{(r)}_{a_{r},b}(x)$~be a Boolean function defined by~(6).~With the help of Boolean functions
\[g_{a_{r},b}(x)=\sum_{ r\in R}\mathrm{Tr}^{m}_{1}  \left  (a^{\prime}_{r}D_{r}  \left   ( \frac{1}{x^{2}+x+\rho_{0}}   \right   )+ a^{\prime\prime}_{r}D_{r}  \left  ( \frac{x^{2}}{ \rho_{0}(x^{2}+x+\rho_{0})}    \right   )   \right  )+\mathrm{Tr}^{2}_{1}(b) ,\]
where~$a^{\prime}_{r}=\frac{a_{r}u^{-r}+a^{2^m}_{r}u^{r}_{0}  }{u^{r}_{0}+u^{-r}_{0} }$~and~$a^{\prime\prime}_{r}=\frac{a_{r}+a^{2^m}_{r}}{u^{r}_{0}+u^{-r}_{0}}$.~A Boolean function~$f^{(r)}_{a_{r},b}(x)$~is said to be a hyper-bent function if and only if
\[ \sum_{ x\in\mathbb{F}_{2^{m}}}(-1)^{g_{a_{r},b}(x)}=1-(-1)^{f^{(r)}_{a_{r},b}(1)}  .\]
\begin{proof}
	Note that the facts
	\[  f^{(r)}_{a_{r},b}(x)=\sum_{ r\in R}\mathrm{Tr}^{n}_{1}(a_{r}x^{r(2^m-1)})+\mathrm{Tr}^{2}_{1}(bx^\frac{2^n-1}{3}),  \]
	and~(1).~A direct verification confirms that~$a^{\prime}_{r},~a^{\prime\prime}_{r}\in\mathbb{F}_{2^{m}}$~and~$a_{r}=a^{\prime}_{r} + a^{\prime\prime}_{r}u^{r}_{0}$,~where the choice of~$u_{0}$~is satisfied for the total functions~$ f^{(r)}_{a_{r},b}(x)$~with their coefficients~$a_{r}$~in the extension field~$\mathbb{F}_{2^{n}}$.~One has
	\begin{align*}
	\Lambda(f^{(r)}_{a_{r},b})~&=\sum_{ u\in U_{2^{m}+1}}(-1)^{\sum\limits_{ r\in R} \mathrm{Tr}^{m}_{1} (a^{\prime}_{r}(u^{r} + u^{-r} ) +a^{\prime\prime}_{r} ((u_{0}u)^{r} +(u_{0}u)^{-r} ) ) + \mathrm{Tr}^{2}_{1}(b)        }
	\end{align*}
	We depend on the facts
	\[   u+\frac{1}{u} =\frac{1}{x^{2}+x+\rho_{0}},~u_{0}u+\frac{1}{u_{0}u} =\frac{x^{2}}{\rho_{0}(x^{2}+x+\rho_{0})} ,                 \]
	and draw support from the properties of Dickson polynomials:
	\[ u^{r}+u^{-r} =D_{r}( u+u^{-1} ),~{(u_{0}u)}^{r}+{(u_{0}u)}^{-r}=D_{r}(  (u_{0}u)+{(u_{0}u)}^{-1}).    \]

\noindent Note that~$U_{2^{m}+1}\setminus{\{1\}}=\left \{ \frac{(u_{0}+1 )x+1} {(u_{0}+1 )x+u_{0}} : x\in\mathbb{F}_{2^{m}}    \right \}$.~Further, we have
\begin{align*}
\Lambda(f^{(r)}_{a_{r},b})~&=\sum_{ u\in U_{2^{m}+1}}  (-1)^{\sum\limits_{ r\in R}\mathrm{Tr}^{m}_{1} (a^{\prime}_{r}D_{r}(u + u^{-1} ) +a^{\prime\prime}_{r}D_{r} ( (u_{0}u) +(u_{0}u)^{-1} ) ) + \mathrm{Tr}^{2}_{1}(b)        }   ~\\
                          ~&=\sum_{ x\in\mathbb{F}_{2^{m}}}    (-1)^{\sum_{ r\in R}\mathrm{Tr}^{m}_{1}  \left  (a^{\prime}_{r}D_{r}  \left   ( \frac{1}{x^{2}+x+\rho_{0}}   \right   )+ a^{\prime\prime}_{r}D_{r}  \left  ( \frac{x^{2}}{ \rho_{0}(x^{2}+x+\rho_{0})}    \right   )   \right  )+\mathrm{Tr}^{2}_{1}(b) } + (-1)^{f^{(r)}_{a_{r},b}(1)}.
\end{align*}
According to~$f^{(r)}_{a_{r},b}(x)$~is hyper-bent function if and only if~$\Lambda(f^{(r)}_{a_{r},b})=1$.~We say that this function~$f^{(r)}_{a_{r},b}(x)$~is hyper-bent function if and only if
	\[  \sum_{ x\in\mathbb{F}_{2^{m}}}    (-1)^{\sum_{ r\in R}\mathrm{Tr}^{m}_{1}  \left  (a^{\prime}_{r}D_{r}  \left   ( \frac{1}{x^{2}+x+\rho_{0}}   \right   )+ a^{\prime\prime}_{r}D_{r}  \left  ( \frac{x^{2}}{ \rho_{0}(x^{2}+x+\rho_{0})}    \right   )   \right  )+\mathrm{Tr}^{2}_{1}(b) } =1-(-1)^{f^{(r)}_{a_{r},b}(1)} .  \]
This is completed the proof.
\end{proof}
\noindent{\bf Theorem 3.2}~~Let~$u_{0}$~be a primitive~$(2^{m}+1)$-th~root of unity.~Let~$ f_{a_{r}}(x)$~be the Boolean function defined on~$\mathbb{F}_{2^{n}}$~by~(5),~where~$a_{r}\in\mathbb{F}_{2^{n}}$~and~$a_{r}=a^{\prime}_{r}+a^{\prime\prime}_{r}u^{r}_{0}$.~Denote that~$g_{a^{\prime}_{r}}(x)=\sum_{ r\in R}\mathrm{Tr}^{m}_{1}(a^{\prime}_{r}D_{r}(x))$~and~$g_{a^{\prime\prime}_{r}}(x)=\sum_{ r\in R}\mathrm{Tr}^{m}_{1}(a^{\prime\prime}_{r}D_{r}(x))$.~We have
\begin{align*}~\tag{7}
\sum\limits_{u\in U_{2^{m}+1}}\chi(f_{a_{r}}(u^{p}))=1+\sum\limits_{u\in U_{2^{m}+1}\setminus \{ 1 \}    }\chi(            g_{a^{\prime}_{r}}(D_{p}( u +u^{-1} ))) + g_{a^{\prime\prime}_{r}}(D_{p}(  u^{\frac{1}{p}}_{0}u   +  (u^{\frac{1}{p}}_{0}u)^{-1}     )))  .
\end{align*}
\begin{proof}
	Let~$u$~be an element of cyclic group~$U_{2^{m}+1}$~and the choice of~$u$~have no impact on the correctness of Theorem 3.2. The formula can be attained as follows:
	\begin{align*}
	f_{a_{r}}(u^{p})=  \sum_{ r\in R } \mathrm{Tr}^{m}_{1} (a^{\prime}_{r}D_{rp} (u +u^{-1} ) + a^{\prime\prime}_{r}D_{rp}     (   u^{\frac{1}{p}}_{0}u   +  (u^{\frac{1}{p}}_{0}u)^{-1} )).
	\end{align*}
	By straightforword calculations,~one has
	\begin{align*}
	\sum\limits_{u\in U_{2^{m}+1}}\chi(f_{a_{r}}(u^{p} ))
	~&=\sum\limits_{u\in U_{2^{m}+1}}\chi(  \sum_{ r\in R } \mathrm{Tr}^{m}_{1} (a^{\prime}_{r}D_{rp} (u +u^{-1} ) + a^{\prime\prime}_{r}D_{rp}     (   u^{\frac{1}{p}}_{0}u   +  (u^{\frac{1}{p}}_{0}u)^{-1} ))    )   ~\\
	~&=  1+  \sum\limits_{u\in U_{2^{m}+1}\setminus \{ 1 \}    }\chi(  \sum_{ r\in R} \mathrm{Tr}^{m}_{1} (a^{\prime}_{r}D_{rp} (u +u^{-1} ) + a^{\prime\prime}_{r}D_{rp}     (   u^{\frac{1}{p}}_{0}u   +  (u^{\frac{1}{p}}_{0}u)^{-1} ))    )~\\
	~&=1+\sum\limits_{u\in U_{2^{m}+1}\setminus \{ 1 \}    }\chi(            g_{a^{\prime}_{r}}(D_{p}( u +u^{-1} ))) + g_{a^{\prime\prime}_{r}}(D_{p}(  u^{\frac{1}{p}}_{0}u   +  (u^{\frac{1}{p}}_{0}u)^{-1}     ))).
	\end{align*}
	This is completed the proof.
\end{proof}
We denote the equation~$T_{1}$~as
\[  \mathrm{T}_{1}(g_{a^{\prime}}(c_{1})+g_{a^{\prime\prime}_{r}}(c_{2}))=  \sum\limits_{u\in U_{2^{m}+1}\setminus \{ 1 \}    }\chi(            g_{a^{\prime}_{r}}( u +u^{-1} ) + g_{a^{\prime\prime}_{r}}( u_{0}u   +  (u_{0}u)^{-1}     )),  \]
where~$c_{1}=u +u^{-1}$~and~$c_{2}= u_{0}u + (u_{0}u)^{-1} $. We employ the fact that Boolean function~$f$~is hyper-bent if and only if~$\sum\limits_{u\in U_{2^{m}+1}}(-1)^{f(u)}=1$, which is leading to the following the corollary.~\\
\noindent{\bf Corollary~3.1}~~Let~$f_{a_{r}}(x)=\sum_{ r\in R} \mathrm{Tr}^{n}_{1} (a_{r}x^{r(2^{m}-1)} )$~be a Boolean function,~where~$a_{r}\in\mathbb{F}_{2^{n}}$.~Then,~$f_{a_{r}}(x)$~is hyper-bent if and only if~$\mathrm{T}_{1}(g_{a^{\prime}}(c_{1})+g_{a^{\prime\prime}_{r}}(c_{2})) =0$.
\begin{proof}
	A Boolean function~$f_{a_{r}}(x)$~is hyper-bent with Dillion exponents defined over~$\mathbb{F}_{2^{m}}$~if and only if
	\[    \sum\limits_{u\in U_{2^{m}+1}}(-1)^{f_{a_{r}}(u)}=1 .  \]
By this equation~$\Lambda(f_{a_{r}}(u))=1+\mathrm{T}_{1}(g_{a^{\prime}}(c_{1})+g_{a^{\prime\prime}_{r}}(c_{2}))  $,~we get the ultimate fact.
\end{proof}
\indent Let~$\alpha$~be a primitive element in~$\mathbb{F}_{2^{n}}$.~If~$U={\{u\in\mathbb{F}^{\ast}_{2^{n}} \mid u^{2^{m}+1}=1  \}}$,~then~$\xi=\alpha^{2^{m}-1}$~be a generator of the cyclic group~$U$~and~$U$~can be decomposed as:~$U=\bigcup^{2}_{i=0}\xi^{i}V$~with~$V=\{u^{3}: u\in U\}$.~Let~$S_{i}$~be the sums~
\[   S_{i}=\sum\limits_{v\in V}\chi(f_{a_{r}}  (\xi^{i}v)), \forall i\in \{0,1,2\} ,  \]
it is an obvious result that~$\sum\limits^{2}_{i=0} S_{i}=\sum\limits_{u\in U}\chi(f_{a_{r}} (u))$.~\\
\noindent{\bf Definition~3.1}~~For~$i\in\mathbb{Z}$,~define~$S_{i}$~and~$\overline{S}_{i}$~to the partial exponential sums
\[ S_{i}=\sum\limits_{ v\in V }\chi(f_{a_{r}}(\zeta^{i}v) ),  \overline{S}_{i}=\sum\limits_{ v\in V }\chi( \overline{f}_{a_{r}}(\zeta^{i}v) ).  \]
\noindent{\bf Remark~3.1}~~It is straightforward to see that
\[    \sum\limits^{2}_{i=0}  S_{i}= \Lambda(f_{a_{r}}(u))=1+\mathrm{T}_{1}(g_{a^{\prime}}(c_{1})+g_{a^{\prime\prime}_{r}}(c_{2})),  \]
thus,~$f_{a_{r}}(x)$~is hyper-bent if and only if~$\sum\limits^{2}_{i=0}  S_{i}=1$.~Furthermore,~we present an equivalent characterization of hyper-bent function as
\[   \mathrm{T}_{1}(g_{a^{\prime}}(c_{1})+g_{a^{\prime\prime}_{r}}(c_{2}))=0.                             \]
\noindent{\bf Proposition~3.1}~~Let~$s$~be the order of subgroup of the field~$\mathbb{F}^{\ast}_{2^{n}}$~and~$ f^{(r)}_{a_{r},b}(x)$~be a Boolean function defined on~$\mathbb{F}_{2^{n}}$~by~(6),~then~$f^{(r)}_{a_{r},b}(x)$~is hyper-bent if and only if~$ \sum\limits^{2}_{i=0} \chi(\mathrm{Tr}^{2}_{1} ( b\xi^{i} ) )  \overline{S}_{i}=1$.
\begin{proof}
	As a matter of fact, we can refer to this way of proving in~[21]~for the value of~$\Lambda(f^{(r)}_{a_{r},b})$.~Since this function~$f^{(r)}_{a_{r},b}(x)$~is hyper-bent if and only if~$\Lambda(f^{(r)}_{a_{r},b})=1$, we get the result that~$f^{(r)}_{a_{r},b}(x)$~is hyper-bent if and only if~$\sum\limits^{2}_{i=0} \chi(\mathrm{Tr}^{2}_{1} ( b\xi^{i} ) )  \overline{S}_{i}(a)=1$.
\end{proof}~\\
\noindent{\bf Theorem~3.3}~~Let~$f_{a_{r}}(x)=\sum_{ r\in R } \mathrm{Tr}^{n}_{1} (a_{r}x^{r(2^{m}-1)} )$~be a Boolean function with~$a_{r}\in\mathbb{F}_{2^{n}}$~and~$\rho_{0}=\frac{u_{0}}{1+u^{2}_{0}}$~with~$u_{0}\in U_{2^{m}+1}\backslash\{1\}$,~then~$f_{a_{r}}(x)$~is said to be a hyper-bent if and only if
\[   wt    \left ( g_{a^{\prime}_{r}}    \left  (\frac{1}{x^{2}+x+\rho_{0}}  \right ) + g_{a^{\prime\prime}_{r}}     \left    (  \frac{x^{2}}{ \rho_{0}(x^{2}+x+\rho_{0})}  \right    )     \right      )= 2^{m-1}  .             \]
\begin{proof}
	It is worthy of attention that both~$\mathrm{Tr}^{n}_{1}( x^{2}+x+\rho_{0}    )=1$~and~$\mathrm{Tr}^{n}_{1}( {(\rho_{0}/x) }^{2}+(\rho_{0}/x)+ \rho_{0}   )=1  $~can be used to determine the hyper-bentness of Boolean function~$f_{a_{r}}(x)$~with~$\mathrm{Tr}^{n}_{1}(\rho_{0})=1$.~Hence,~we get the results
	\begin{align*}
	\sum\limits_{u\in U_{2^{m}+1}}\chi(f_{a_{r}}(u ))~&=1+\sum\limits_{u\in U_{2^{m}+1}\setminus \{ 1 \}    }\chi(            g_{a^{\prime}_{r}}( u +u^{-1} ) + g_{a^{\prime\prime}_{r}}( u_{0}u   +  (u_{0}u)^{-1} ))~\\
	~&=1+\sum\limits_{x\in\mathbb{F}_{2^{m} } }\chi  \left    (   g_{a^{\prime}_{r}}\left (\frac{1}{x^{2}+x+\rho_{0}}  \right   ) + g_{a^{\prime\prime}_{r}}\left (  \frac{x^{2}}{ \rho_{0}(x^{2}+x+\rho_{0})}\right  )   \right   )~\\
	~&=1+2\sum\limits_{      x\in\mathbb{F}_{2^{m}},~\mathrm{Tr}^{m}_{1} (1/x)=1       }\chi(            g_{a^{\prime}_{r}}(x) + g_{a^{\prime\prime}_{r}}(x)  ).
	\end{align*}
	For a Boolean function~$h$~defined on~$\mathbb{F}_{2^{n}}$, it is~satisfied with the equation~$\sum\limits_{x\in\mathbb{F}_{2^{n} } }\chi(h(x))=2^{n}-2wt(h)$.~Thus,
	\[\sum\limits_{x\in\mathbb{F}_{2^{m} } }\chi \left (  g_{a^{\prime}_{r}} \left (\frac{1}{x^{2}+x+\rho_{0}}  \right  ) + g_{a^{\prime\prime}_{r}} \left   (  \frac{x^{2}}{ \rho_{0}(x^{2}+x+\rho_{0})}  \right    ) \right  )   =2^{m}-2wt \left  ( g_{a^{\prime}_{r}} \left (\frac{1}{x^{2}+x+\rho_{0}} \right  ) + g_{a^{\prime\prime}_{r}}   \left (  \frac{x^{2}}{ \rho_{0}(x^{2}+x+\rho_{0})}\right  )   \right    )  .    \]
	By Corollary 3.1,~$f_{a_{r}}(x)$~is hyper-bent if and only if
	\[      \sum\limits_{      x\in\mathbb{F}_{2^{m}},~\mathrm{Tr}^{m}_{1} (1/x)=1       }\chi(            g_{a^{\prime}_{r}}(x) + g_{a^{\prime\prime}_{r}}(x))  = 0.   \]
This implies that
	\[  wt \left  ( g_{a^{\prime}_{r}} \left (\frac{1}{x^{2}+x+\rho_{0}} \right  ) + g_{a^{\prime\prime}_{r}}   \left (  \frac{x^{2}}{ \rho_{0}(x^{2}+x+\rho_{0})}\right  )   \right    )  = 2^{m-1}  .             \]
\end{proof}
\noindent{\bf Remark 3.2}~~If~$a_{r}\in\mathbb{F}_{2^{m}}$, then~$a^{\prime}_{r}=a_{r}$,~$a^{\prime\prime}_{r}=0$~and~$f^{(r)}_{a_{r},b}(1)=\mathrm{Tr}^{2}_{1}(b)$. Under the conditions,~$f^{(r)}_{a_{r},b}(x)$~is said to be a hyper-bent function if and only if
\[ \sum_{ x\in\mathbb{F}_{2^{m}}}(-1)^{    \sum\limits_{ r\in R}\mathrm{Tr}^{m}_{1} (a_{r}D_{r}( \frac{1}{x^{2}+x+\rho_{0}} ) ) } =1-(-1)^{\mathrm{Tr}^{2}_{1}(b)} .\]
\noindent{\bf Theorem 3.4}~~Let~$n=2m$~and~$a_{r}\in\mathbb{F}_{2^{n}}$,~$u_{0}$~be a primitive~$(2^{m}+1)$-th~root of unity and~$\rho_{0}=\frac{u_{0}}{1+u^{2}_{0}}$.~Let~$g_{a_{r},1}(x)$~be a Boolean function defined on~$\mathbb{F}_{2^{m}}$~by
\[g_{a_{r},1}(x)=\sum_{ r\in R}\mathrm{Tr}^{m}_{1} \left (a^{\prime}_{r}D_{r}   \left   ( \frac{1}{x^{2}+x+\rho_{0}}   \right  )+ a^{\prime\prime}_{r}D_{r}    \left   ( \frac{x^{2}}{ \rho_{0}(x^{2}+x+\rho_{0})}    \right    )\right ) ,\]
where~$a^{\prime}_{r}=\frac{a_{r}u^{-r}_{0}+a^{2^m}_{r}u^{r}_{0}  }{u^{r}_{0}+u^{-r}_{0} }$~and~$a^{\prime\prime}_{r}=\frac{a_{r}+a^{2^m}_{r}}{u^{r}_{0}+u^{-r}_{0}}$.~Then,~$f^{(r)}_{a_{r},1}(x)$~is hyper-bent function if and only if
\[ \sum_{ x\in\mathbb{F}_{2^{m}}}(-1)^{g_{a_{r},1}(x)}=1-(-1)^{f^{(r)}_{a_{r},1}(1)}  .\]
\begin{proof}
	Let~$\Lambda(f^{(r)}_{a_{r},1})=\sum_{ u\in U_{2^{m}+1}}(-1)^{ f^{(r)}_{a_{r},1}(u)       }    $. Applying~this relation~$a_{r}=a^{\prime}_{r} + a^{\prime\prime}_{r}u^{r}_{0}$,~one has
	\begin{align*}
	\Lambda(f^{(r)}_{a_{r},1})~&=\sum_{ u\in U_{2^{m}+1}}(-1)^{\sum\limits_{ r\in R} \mathrm{Tr}^{m}_{1} (a^{\prime}_{r}(u^{r} + u^{-r} ) +a^{\prime\prime}_{r} ((u_{0}u)^{r} +(u_{0}u)^{-r} ) )       } ~\\
	~&=\sum_{ u\in U_{2^{m}+1}}  (-1)^{\sum\limits_{ r\in R}\mathrm{Tr}^{m}_{1} (a^{\prime}_{r}D_{r}(u + u^{-1} ) +a^{\prime\prime}_{r} D_{r}( (u_{0}u) +(u_{0}u)^{-1} ) )       }   ~\\
	~&=\sum_{ x\in\mathbb{F}_{2^{m}}}    (-1)^{\sum\limits_{ r\in R}\mathrm{Tr}^{m}_{1} (a^{\prime}_{r}D_{r}(u + u^{-1} ) +a^{\prime\prime}_{r} D_{r}( (u_{0}u) +(u_{0}u)^{-1} )   ) } + (-1)^{f^{(r)}_{a_{r},1}(1)},
	\end{align*}
	where~$a^{\prime}_{r},~a^{\prime\prime}_{r}\in\mathbb{F}_{2^{m}}$. Then, Boolean function~$f^{(r)}_{a_{r},1}(x)$~is hyper-bent function if and only if
	\[  \sum_{ x\in\mathbb{F}_{2^{m}}}    (-1)^{ \sum\limits_{ r\in R}\mathrm{Tr}^{m}_{1} (a^{\prime}_{r}D_{r}(u + u^{-1} ) +a^{\prime\prime}_{r}D_{r} ( (u_{0}u) +(u_{0}u)^{-1} ) ) } =1-(-1)^{f^{(r)}_{a_{r},1}(1)} .  \]
	This is completed the proof.
\end{proof}
\indent~Let us investigate this Boolean function~$f_{a,b,c}(x)$~together. We define
\begin{align*}~\tag{8}
f_{a,b,c}(x)=\mathrm{Tr}^{n}_{1}((a+cu_{0})x^{(2^{m}-1)}+  bx^{\frac{ 2^{n}-1}{3}}   ),~~a,b,c\in\mathbb{F}_{2^{m}},
\end{align*}
where~$m$~is an odd integer and~$u_{0}\in\mathbb{F}_{4}\setminus{\mathbb{F}_{2}}$.~Then, we get that
\begin{align*}
g(x)~&=\mathrm{Tr}^{m}_{1}\left ( aD_{1}\left (  \frac{1}{x^{2}+x+1}  \right  )    +   cD_{1} \left (  \frac{x^{2}}{x^{2}+x+1} \right  )   \right  )~\\
~&=\mathrm{Tr}^{m}_{1}\left (  \frac{cx^{2}+a}{x^{2}+x+1}        \right  ).
\end{align*}
\noindent{\bf Theorem 3.5}~~Let~$ f_{a,b,c}(x)$~be defined by~(8) with~$u_{0}\in U_{2^{m}+1}\setminus{\{1\}}$,~where~$m$~is an odd integer. Then,~$ f_{a,b,c}(x)$~is said to be a hyper-bent function if and only if
\[  \sum_{ x\in\mathbb{F}_{2^{m}}}    (-1)^{  \mathrm{Tr}^{m}_{1}\left (  \frac{c x^{2}+a}{x^{2}+x+1} \right  )   } =1-(-1)^{ \mathrm{Tr}^{m}_{1}(c)} .  \]
\begin{proof}
Since~$u_{0}$~is a primitive third root of unity, this equation~$u_{0}+\frac{1}{u_{0}}=1$~can be obtained from~$\rho_{0}=\frac{u_{0}}{1+u^{2}_{0}}$. It is easy to get~$\Lambda(f_{a,b,c})$~from the proof process of Theorem 3.1.
\begin{align*}
\Lambda(f_{a,b,c})~&=\sum_{ x\in\mathbb{F}_{2^{m}}}    (-1)^{g\left(x \right) }+\left(-1 \right)^{f_{a,b,c}(x)} ~\\
                  ~&= \sum_{ x\in\mathbb{F}_{2^{m}}}    (-1)^{  \mathrm{Tr}^{m}_{1}\left (  \frac{c x^{2}+a}{x^{2}+x+1} \right  )   } +(-1)^{ \mathrm{Tr}^{m}_{1}(c)} .
\end{align*}
Therefore,~by Theorem 3.1,~this function~$f_{a,b,c}(x)$~is a hyper-bent function if and only if
\[  \sum_{ x\in\mathbb{F}_{2^{m}}}    (-1)^{ g\left(x \right)   } =1-(-1)^{ \mathrm{Tr}^{m}_{1}(c)} .  \]
This is completed the proof.
\end{proof}

\indent~We have the following result that describes the hyper-bentness of this function~$ f_{a,b,c}(x)$~with the hyperelliptic curves.~\\
\noindent{\bf Theorem 3.6}~~A Boolean function~$ f_{a,b,c}(x)$~defined by~(8).~The curve~$C$~is written as
\[ \mathcal{C}:~~y^{2}+yD(x)=cx^{2}D(x)+aD(x),                  \]
where~$D(x)=x^{2}+x+1$.~This function~$ f_{a,b,c}(x)$~is hyper-bent function if and only if
\[   \# \mathcal{C} (\mathbb{F}_{2^{m}})=2^{m}+2-(-1)^{ \mathrm{Tr}^{m}_{1}(c)} .  \]
\begin{proof}
	Let the sum
	\begin{align*}
	L~&=1-(-1)^{ \mathrm{Tr}^{m}_{1}(c)} ~\\
	~&=\sum_{ x\in\mathbb{F}_{2^{m}}}    (-1)^{  \mathrm{Tr}^{m}_{1}\left (  \frac{c x^{2}+a}{x^{2}+x+1} \right  )   }.
	\end{align*}
	Denote~$T$~is the number of~$x\in\mathbb{F}_{2^{m}}$~such that~$\mathrm{Tr}^{m}_{1}\left (  \frac{c x^{2}+a}{x^{2}+x+1} \right  )=0 $, then we have
	\[   2T=  \# \mathcal{C}_{1} (\mathbb{F}_{2^{m}})-1   .   \]
	The curve~$\mathcal{C}_{1}$~is denoted as
	\begin{align*}~\tag{9}
	t^{2}+t=  \frac{c x^{2}+a}{x^{2}+x+1} .
	\end{align*}
	The substitution~$y=t(x^{2} +x +1  )$~into~(9)~is a bijective transformation between~$\mathcal{C}_{1}(\mathbb{F}_{2^{m}})$~and~$\mathcal{C}(\mathbb{F}_{2^{m}})$.~Furthermore,~one has
	\[   L = \# \mathcal{C} (\mathbb{F}_{2^{m}})-1-2^{m}   .                  \]
	\indent We characterize the hyper-bentness of this function~$ f_{a,b,c}(x)$~with hyperelliptic curves in terms of Theorem 3.5,~i.e.,~this function~$ f_{a,b,c}(x)$~is hyper-bent if and only if~$\# \mathcal{C} (\mathbb{F}_{2^{m}})=2^{m}+2-(-1)^{ \mathrm{Tr}^{m}_{1}(c)}  $.~This proof is completed.
\end{proof}~\\
\noindent{\bf Example 3.1}~~To illustrate our results,~Let~$\rho_{0}=\frac{u_{0}}{1+u^{2}_{0}}$,~$u_{0}\in U_{2^{m}+1}\setminus{\{1\}}$~and~$ f^{(1)}_{a_{1},b}(x)=\mathrm{Tr}^{n}_{1} (a_{1}x^{2^{m}-1})+\mathrm{Tr}^{2}_{1}(bx^{  \frac{2^{n}-1}{3} })$~be a Boolean function defined on~$\mathbb{F}_{2^{n}}$~with~$a_{1}=a^{\prime}_{1}+a^{\prime\prime}_{1}u^{r}_{0}$,~where~$a^{\prime}_{1}, a^{\prime\prime}_{1}\in\mathbb{F}_{2^{m}}$.~Let~$g_{a_{1},b}$~be a Boolean function defined on~$\mathbb{F}_{2^{m}}$~by
\[g_{a_{1},b}(x)=\mathrm{Tr}^{m}_{1} \left (a^{\prime}_{1} \frac{1}{x^{2}+x+\rho_{0}} + a^{\prime\prime}_{1} \frac{x^{2}}{ \rho_{0}(x^{2}+x+\rho_{0})} \right )+\mathrm{Tr}^{2}_{1}(b).\]
Then, by Theorem 3.1,~$f^{(1)}_{a_{1},b}$~is hyper-bent if and only if
\[  \sum_{ x\in\mathbb{F}_{2^{m}}}    (-1)^{g_{a_{1},b}(x)} =1-(-1)^{f^{(r)}_{a_{1},b}(1)} . \]
\noindent{\bf Example 3.2}~~Let us consider a particular~Boolean function~$f^{(r)}_{a_{r},1}(x)$~defined on~$\mathbb{F}_{2^{n}}$~as
\[ f^{(r)}_{a_{r},1}(x)=\mathrm{Tr}^{n}_{1}(a_{r} x^{r(2^m-1)} )+\mathrm{Tr}^{2}_{1}(x^\frac{2^n-1}{3}),~\forall x\in \mathbb{F}_{2^{n}},         \]
where~$a_{r}\in\mathbb{F}_{2^{n}}$.~Let~$u_{0}$~be a primitive~$(2^{m}+1)$-th~root of unity and~$\rho_{0}=\frac{u_{0}}{1+u^{2}_{0}}$,~$g_{a_{r},1}$~be a Boolean functions defined on~$F_{2^{m}}$~by
\[g_{a_{r},1}(x)=\mathrm{Tr}^{m}_{1}  \left  (a^{\prime}_{r}D_{r} \left  ( \frac{1}{x^{2}+x+\rho_{0}}  \right  )+ a^{\prime\prime}_{r}D_{r} \left  ( \frac{x^{2}}{ \rho_{0}(x^{2}+x+\rho_{0})} \right )  \right ) ,\]
where~$a^{\prime}_{r}=\frac{a_{r}u^{-r}+a^{2^m}_{r}u^{r}_{0}  }{u^{r}_{0}+u^{-r}_{0} }$~and~$a^{\prime\prime}_{r}=\frac{a_{r}+a^{2^m}_{r}}{u^{r}_{0}+u^{-r}_{0}}$.~Then, by Theorem 3.4,~$f^{(r)}_{a_{r},1}(x)$~is hyper-bent function if and only if
\[ \sum_{ x\in\mathbb{F}_{2^{m}}}(-1)^{g_{a_{r},1}(x)}=1-(-1)^{f^{(r)}_{a_{r},1}(1)}  .\]
\section{Conclusion}\label{4}
\indent~~We answer an open problem left in~[34]~that determines hyper-bentness of Charpin-Gong like functions~$f^{(r)}_{a_{r},b}(x)=\sum\limits_{ r\in R}\mathrm{Tr}^{n}_{1}(a_{r}x^{r(2^m-1)})+\mathrm{Tr}^{2}_{1}(bx^\frac{2^{n}-1}{3})$,~where the coefficient~$b$~is in~$\mathbb{F}_{4}$,~some of the coefficients~$a_{r}$~are in~$\mathbb{F}_{2^{n}}$,~but not in~$\mathbb{F}_{2^{m}}$.~In this paper,~we provide several characterizations of hyper-bentness for Charpin-Gong like functions~$f^{(r)}_{a_{r},b}(x)$ with the oefficient~$a_{r}$~is in~$\mathbb{F}_{2^{n}}$~by applying the technology in~[37].~A relative problem is to character the hyper-bentness of this family of Boolean function~$f^{(r)}_{a_{r},b}(x)=\sum\limits_{ r\in R}\mathrm{Tr}^{n}_{1}(a_{r}x^{r(2^m-1)})+\mathrm{Tr}^{4}_{1}(bx^\frac{2^n-1}{5})$,~where the same restriction~$a_{r}$~is in~$\mathbb{F}_{2^{n}}$~rather than in~$\mathbb{F}_{2^{m}}$~and satisfies~$m\equiv 2(mod4)$~and the coefficient~$b$ is in~$\mathbb{F}_{16}$.~In additional,~to generalize the results of this paper to~$p$-ary hyper-bent functions is also a fascinating problem.


\bibliographystyle{unsrt}
\bibliography{cas-model2-names}

\bio{}
[1] Oscar S Rothaus. On “bent” functions. Journal of Combinatorial Theory, Series A, 20(3):300–305, 1976.
\endbio
\bio{}
[2] Anne Canteaut, Claude Carlet, Pascale Charpin, and Caroline Fontaine. On cryptographic properties of the cosets of r (1, m). IEEE Transactions on Information Theory, 47(4):1494–1513, 2001.
\endbio
\bio{}
[3] Florence Jessie MacWilliams and Neil James Alexander Sloane. The theory of error-correcting codes, volume 16. Elsevier, 1977.
\endbio
\bio{}
[4] Claude Carlet, Yves Crama, and Peter L Hammer. Boolean functions for cryptography and error-correcting codes., 2010.
\endbio
\bio{}
[5] John Olsen, Robert Scholtz, and Lloyd Welch. Bent-function sequences. IEEE Transactions on Information Theory, 28(6):858–864, 1982.
\endbio
\bio{}
[6] Yin Tan, Alexander Pott, and Tao Feng. Strongly regular graphs associated with ternary bent functions. Journal of Combinatorial Theory,
Series A, 117(6):668–682, 2010.
\endbio
\bio{}
[7] Yeow Meng Chee, Yin Tan, and Xian De Zhang. Strongly regular graphs constructed from p-ary bent functions. Journal of Algebraic
Combinatorics, 34:251–266, 2011.
\endbio
\bio{}
[8] Alexander Pott, Yin Tan, Tao Feng, and San Ling. Association schemes arising from bent functions. Designs, Codes and Cryptography,
59:319–331, 2011.
\endbio
\bio{}
[9] Anne Canteaut, Pascale Charpin, and Gohar M Kyureghyan. A new class of monomial bent functions. Finite Fields and Their
Applications, 14(1):221–241, 2008.
\endbio
\bio{}
[10] Claude Carlet. Two new classes of bent functions. In Advances in Cryptology—EUROCRYPT’93: Workshop on the Theory and
Application of Cryptographic Techniques Lofthus, Norway, May 23–27, 1993 Proceedings 12, pages 77–101. Springer, 1994.
\endbio
\bio{}
[11] Claude Carlet and Sihem Mesnager. On dillons class h of bent functions, niho bent functions and o-polynomials. Journal of Combinatorial
Theory, Series A, 118(8):2392–2410, 2011.
\endbio
\bio{}
[12] Pascale Charpin, Enes Pasalic, and Cédric Tavernier. On bent and semi-bent quadratic boolean functions. IEEE Transactions on
Information Theory, 51(12):4286–4298, 2005.
\endbio
\bio{}
[13] John Francis Dillon. ELEMENTARY HADAMARD DIFFERENCE-SETS. University of Maryland, College Park, 1974.
\endbio
\bio{}
[14] Tor Helleseth and Alexander Kholosha. New binomial bent functions over the finite fields of odd characteristic. In 2010 IEEE
International Symposium on Information Theory, pages 1277–1281. IEEE, 2010.
\endbio
\bio{}
[15] Wenjie Jia, Xiangyong Zeng, Tor Helleseth, and Chunlei Li. A class of binomial bent functions over the finite fields of odd characteristic.
IEEE transactions on information theory, 58(9):6054–6063, 2012.
\endbio
\bio{}
[16] Khoongming Khoo, Guang Gong, and Douglas R Stinson. A new characterization of semi-bent and bent functions on finite fields.
Designs, Codes and Cryptography, 38:279–295, 2006.
\endbio
\bio{}
[17] Shenghua Li, Lei Hu, and Xiangyong Zeng. Constructions of p-ary quadratic bent functions. Acta Applicandae Mathematicae,
100(3):227–245, 2008.
\endbio
\bio{}
[18] Nian Li, Tor Helleseth, Xiaohu Tang, and Alexander Kholosha. Several new classes of bent functions from dillon exponents. IEEE
transactions on information theory, 59(3):1818–1831, 2012.
\endbio
\bio{}
[19] S-C Liu and John J Komo. Nonbinary kasami sequences over gf (p). IEEE transactions on information theory, 38(4):1409–1412, 1992.
\endbio
\bio{}
[20] Sihem Mesnager. Bent functions, volume 1. Springer, 2016.
\endbio
\bio{}
[21] Sihem Mesnager and Jean-Pierre Flori. Hyperbent functions via dillon-like exponents. IEEE transactions on information theory,
59(5):3215–3232, 2013.
\endbio
\bio{}
[22] Nam Yul Yu and Guang Gong. Constructions of quadratic bent functions in polynomial forms. IEEE Transactions on Information Theory,
52(7):3291–3299, 2006.
\endbio
\bio{}
[23] Amr M Youssef and Guang Gong. Hyper-bent functions. In Advances in Cryptology—EUROCRYPT 2001: International Conference on
the Theory and Application of Cryptographic Techniques Innsbruck, Austria, May 6–10, 2001 Proceedings 20, pages 406–419. Springer,
2001.
\endbio
\bio{}
[24] Guang Gong and Solomon W Golomb. Transform domain analysis of des. IEEE transactions on Information Theory, 45(6):2065–2073,
1999.
\endbio
\bio{}
[25] Pascale Charpin and Guang Gong. Hyperbent functions, kloosterman sums, and dickson polynomials. IEEE Transactions on Information
Theory, 54(9):4230–4238, 2008.
\endbio
\bio{}
[26] Claude Carlet and Philippe Gaborit. Hyper-bent functions and cyclic codes. Journal of Combinatorial Theory, Series A, 113(3):466–482,
2006.
\endbio
\bio{}
[27] Nils Gregor Leander. Monomial bent functions. IEEE Transactions on Information Theory, 52(2):738–743, 2006.
\endbio
\bio{}
[28] Gilles Lachaud and Jacques Wolfmann. The weights of the orthogonals of the extended quadratic binary goppa codes. IEEE transactions
on information theory, 36(3):686–692, 1990.
\endbio
\bio{}
[29] Sihem Mesnager. A new class of bent and hyper-bent boolean functions in polynomial forms. Designs, Codes and Cryptography,
59(1):265–279, 2011.
\endbio
\bio{}
[30] Sihem Mesnager. A new family of hyper-bent boolean functions in polynomial form. In IMA International Conference on Cryptography
and Coding, pages 402–417. Springer, 2009.
\endbio
\bio{}
[31] Sihem Mesnager. A new class of bent boolean functions in polynomial forms. In Proceedings of International Workshop on Coding and
Cryptography, WCC, pages 5–18, 2009.
\endbio
\bio{}
[32] Chunming Tang, Peng Han, Qi Wang, Jun Zhang, and Yanfeng Qi. The solution to an open problem on the bentness of mesnager’s
functions. Finite Fields and Their Applications, 88:102170, 2023.
\endbio
\bio{}
[33] Sihem Mesnager. Bent and hyper-bent functions in polynomial form and their link with some exponential sums and dickson polynomials.
IEEE transactions on information theory, 57(9):5996–6009, 2011.
\endbio
\bio{}
[34] Sihem Mesnager. Hyper-bent boolean functions with multiple trace terms. In International Workshop on the Arithmetic of Finite Fields,
pages 97–113. Springer, 2010.
\endbio
\bio{}
[35] Chunming Tang, Yanfeng Qi, Maozhi Xu, Baocheng Wang, and Yixian Yang. A new class of hyper-bent boolean functions with multiple
trace terms. Cryptology ePrint Archive, 2011.
\endbio
\bio{}
[36] Cunsheng Ding. A construction of binary linear codes from boolean functions. Discrete mathematics, 339(9):2288–2303, 2016.
\endbio
\bio{}
[37] Claude Carlet, Peng Han, Sihem Mesnager, and Chunming Tang. Möbius transformations and characterizations of hyper-bent functions
from dillon-like exponents with coefficients in extension fields. Advances in Mathematics of Communications, 16(4), 2022.
\endbio


\end{document}